\documentclass[]{article}
\usepackage[]{amsmath}
\usepackage{amssymb}

\numberwithin{equation}{section}
\newtheorem{theorem}[equation]{Theorem}
\newtheorem{corollary}[equation]{Corollary}

\newtheorem{proposition}[equation]{Proposition}
\newtheorem{definition}[equation]{Definition}

\begin{document}
\title{Convexity and the Separability Problem of Quantum Mechanical Density Matrices}
\author{Arthur O. Pittenger 
\thanks {Part of the research for this paper was done at the Centre
for Quantum Computation at Oxford, and their hospitality is gratefully
acknowledged. } \\
Department of Mathematics and Statistics\\
University of Maryland, Baltimore County\\
Baltimore, MD 21250
\and
 Morton H. Rubin 
 \thanks {Support for this work was 
 provided by the Office of Naval
Research and ARDA-NSA }\\
Department of Physics\\
University of Maryland, Baltimore County\\
Baltimore, MD 21250}
\date{9 March 2001 }
\maketitle

\begin{abstract}
A finite dimensional quantum mechanical system is modeled by a \textit{%
density} $\rho $, a trace one, positive semi-definite matrix on a suitable
tensor product space $H^{\left[ N\right] }$. For the system to demonstrate
experimentally certain non-classical behavior, $\rho $ cannot be in $S$, a
closed convex set of densities whose extreme points have a specificed tensor
product form. Two mathematical problems in the quantum computing literature
arise from this context: (1) the determination whether a given $\rho $ is in 
$S$ and (2) a measure of the ``entanglement'' of such a $\rho $ in terms of
its distance from $S$. In this paper we describe these two problems in
detail for a linear algebra audience, discuss some recent results from the
quantum computing literature, and prove some new results.We emphasize the
roles of densities $\rho $ as both operators on the Hilbert space $H^{\left[
N\right] }$ and also as points in a real Hilbert space $M$. We are able to
compute the nearest separable densities $\tau _{0}$ to $\rho _{0}$ in
particular classes of inseparable densities and we use the Euclidean
distance between the two in $M$ to quantify the entanglement of $\rho _{0}$.
We also show the role of $\tau _{0}$ in the construction of separating
hyperplanes, so-called entanglement witnesses in the quantum computing
literature.
\end{abstract}

\section{\textbf{Introduction}}

The idea of using quantum mechanical systems as computing devices arose
during the early 1980s, and examples of the theoretical efficacy of such
devices were soon developed. However, the subject remained primarily a topic
in the theoretical computer science and physics communities until 1994 when
Peter Shor published a \textit{quantum} algorithm for factoring a large
composite integer $N$. Since his algorithm was polynomial rather than
exponential in the number of digits of $N$, it showed that a prospective
quantum computer could factor $N$ more efficiently than was (or is) known to
be possible on a classical computer. As a result quantum versions of
algorithms, information theory and computational complexity have became
subjects of widespread theoretical study, and efforts to actually construct
physical systems which could serve as components of a quantum computer have
become a recognized and active part of experimental physics.

By its very nature, the field of quantum computation and quantum information
theory is highly interdisciplinary and intersects with a variety of
subspecialities in mathematics, computer science, physics and even in
philosophy. The purpose of this paper is to describe one particular problem
in the field of quantum computation which should be of particular interest
to the linear algebra community, and the rest of the paper is devoted to a
mathematical overview of this topic and to the presentation of some new
results. For the reader who would like more background in the subject of
quantum computation, \cite{SL} is an early survey article while \cite
{NC,pitt,Pr} contain descriptions of the subject and additional references.
Other introductory material can be found on various web sites such as that
maintained by the Centre for Quantum Computation at Oxford University \cite
{CQ}.

In the next section we give the mathematical notation necessary to describe
the separability problem, which is related to the physical problem of
constructing a system that produces non-classical phenomena. Essentially,
the mathematical context is one of two nested compact convex sets and the
determination whether a point in the larger set is in the smaller set. In
section 3 we briefly describe the issue of quantifying non-separability or
``entanglement'' and settle on a measure which is the Euclidean distance of
a point to the smaller convex set. In section 4 we present some basic
topological results related to the separability problem, and in section 5 we
develop the role of orthogonality in the analysis. Section 6 deals with
separating hyperplanes, called entanglement witnesses in the quantum
computing literature, and relates them to the earlier analysis. The last two
sections deal with very specific situations in which all of the computations
can be carried out explicitly and which were motivated by basic examples in
the quantum mechanics literature.

For those familiar with related work in the quantum computing community, we
have emphasized convexity and the geometry of the underlying Hilbert space
to provide a useful perspective of the separability problem and related
topics such as entanglement witnesses. We have also shown how the resulting
geometric insight facilitates the extension of results in \cite{witru} as
well as the explicit computation of the nearest separable density to certain
given inseparable densities.

\section{\textbf{Notation and the separability problem}}

Here is the context. Let $H^{\left( N\right) }$ denote an $N=d_{1}\times
\cdots \times d_{n}$ dimensional complex Hilbert space defined as the tensor
product $H^{\left( d_{1}\right) }\otimes \cdots \otimes H^{\left(
d_{n}\right) }$. $\mathit{M}$ is the real Hilbert space of $N\times N$
Hermitian matrices over $H^{\left( N\right) }$ with a real inner product
defined by 
\begin{equation}
\left\langle \left\langle A,B\right\rangle \right\rangle =Tr\left(
A^{\dagger }B\right) =\sum_{j,k}\left( a_{jk}^{\dagger }b_{kj}\right)
=\sum_{j,k}\left( a_{jk}b_{kj}\right)
\end{equation}
which is independent of the particular orthogonal basis of $H^{\left(
N\right) }$ used to define the matrix elements. $\mathit{D}$ denotes the
compact, convex subset of densities; that is, $\rho $ in $\mathit{D}$ is a
positive semidefinite, trace one, $N\times N$ Hermitian matrix which can be
interpreted as the \textit{state} of an $n$-particle system where the $k$'th
particle has $d_{k}$ levels. The \textit{separable} \textit{states}
(densities) comprise a compact convex subset $\mathit{S}$ of $\mathit{D}$,
and $\mathit{S}$ is defined as the closed convex hull of the separable
projections $\otimes _{k}\left| \psi _{k}\right\rangle \left\langle \psi
_{k}\right| $.

In this paper we will consistently use Dirac notation, so that a \textit{ket}
$\left| \psi _{k}\right\rangle $ denotes a column vector in the $d_{k}$
dimensional Hilbert space $H^{\left( d_{k}\right) }$ and the \textit{bra }$%
\left\langle \psi _{k}\right| $ is a $d_{k}$-long row vector whose entries
are the complex conjugates of those of $\left| \psi _{k}\right\rangle $. The
outer product $\left| \psi _{k}\right\rangle \left\langle \psi _{k}\right| $
is a rank $1$, $d_{k}\times d_{k}$ matrix, and the inner product of $\left|
\psi \right\rangle $ and $\left\langle \varphi \right| $ is denoted by the
bracket $\left\langle \varphi |\psi \right\rangle $. In the physics
literature the term \textit{pure} \textit{state} is sometimes used for both
the rank one density $\left| \psi \right\rangle \left\langle \psi \right| $
and the ket $\left| \psi \right\rangle $. Usually the meaning is clear from
the context. (For an introduction to this notation in the context of quantum
computing, see for example \cite{pitt}.)

It follows that the densities in $\mathit{D}$ and $\mathit{S}$ are both
operators on the Hilbert space $H^{\left( N\right) }$ and also points in a
closed convex set in a real Hilbert space $M$. It is that dual role which
underlies our analysis.

When the system of n particles is modelled by densities not in $\mathit{S}$,
some striking quantum effects can be observed. Thus, physical experiments
need to be designed so that the resulting density is in $\mathit{D}-\mathit{S%
}=\mathit{D}\cap \mathit{S}^{c}$. The related \textit{separability problem}
is the mathematical question of how to determine if a given density $\rho $
is in $\mathit{S}$.

This is not an easy question to answer in this generality, and a simple
example illustrates the difficulty. Consider a system with two 2-level
particles, that is to say two ``quantum bits'' or \textit{qubits}, so that $%
N=4$ and $H^{\left( 4\right) }=H^{\left( 2\right) }\otimes H^{\left(
2\right) }$. For this example choose $\rho _{0}=\left| \psi
_{0}\right\rangle \left\langle \psi _{0}\right| $ where $\left| \psi
_{0}\right\rangle =\frac{1}{\sqrt{2}}\left( \left| 00\right\rangle +\left|
11\right\rangle \right) $ and we are using the usual binary notation for two
level systems. Thus, $\rho _{0}$ is a $4\times 4$ matrix with $1/2$ in the
four corners and $0$'s elsewhere.

Define $\rho \left( s\right) =\left( 1-s\right) D_{0}+s\rho _{0}$, where
here and throughout the paper we let the density $D_{0}$ denote the
``normalized'' identity of suitable dimension, $\frac{1}{N}I$. It is easy to
show that $\rho \left( 0\right) =D_{0}$ is in $S$ and, since $\rho _{0}$ is
a projection and thus an extreme point in $\mathit{D}$, that $\rho \left(
1\right) =\rho _{0}$ is not in $\mathit{S}$. Thus there is an intermediate
value $s_{0}$ such that $\rho \left( s_{0}\right) $ is in $\mathit{S}$ and $%
\rho \left( s\right) \notin \mathit{S}$ for $s_{0}<s\leq 1$.

Peres (\cite{peres1}) has observed that a necessary condition for a density
to be separable is that its partial transposes are densities, where the $s$%
'th partial transpose of a density $\rho $ in a given basis respecting the
tensor product is defined by 
\begin{equation}
\rho ^{t_{s}}\left( j_{1}\ldots j_{s}\ldots j_{n},k_{1}\ldots k_{s}\ldots
k_{n}\right) =\rho \left( j_{1}\ldots k_{s}\ldots j_{n},k_{1}\ldots
j_{s}\ldots k_{n}\right) \text{.}
\end{equation}
(Technically we probably should call this part of a generalized Peres
condition, but that seems a bit fussy.) The necessity of the Peres condition
is easy to confirm. If $\rho _{s}$ is a trace one, positive semidefinite
matrix on $H^{\left( d_{s}\right) }$, then so is its (ordinary) transpose $%
\rho _{s}^{t}.$ It then follows that the $s$'th partial transform of the
separable density $\rho _{1}\otimes \ldots \otimes \rho _{n}$ is separable,
and thus if $\rho $ is a convex combination of separable densities, $\rho
=\sum_{a}p_{a}\rho _{a}$, the $s$'th partial transpose of $\rho $ is also in 
$\mathit{S}$. In fact for the $2\times 2$ and $2\times 3$ tensor product
cases the Peres condition is also sufficient (\cite{horo1}). Using that
result it easy to see that for the two qubit example $s_{0}=1/3$, and in
fact it is possible to find an explicit separable representation of $\rho
\left( 1/3\right) $, as noted below and in \cite{pitrub1} for example.
(Related and earlier references include \cite{braun,caves1,werner1,zycz}.)
We discuss this example in more detail below, but suffice it to say here
that the eigenvalues of $\rho \left( 1/3\right) $ are strictly positive so
that $\rho \left( 1/3\right) $ is in the relative interior of $D$.

\section{\textbf{Measures of entanglement}}

A second theme of recent research has been to find a way to quantify the
non-separability or \textit{entanglement} of a system with density $\rho $.
There has also been extensive work in this area, and some representative
papers describing various approaches and basic properties which an
entanglement measure should possess include \cite{benn1,benn2,
karlew,vidal1,vedral2,plenvid,vidal2} among others. The motivation for such
a measure is the recognition that entanglement constitutes a resource which
can be used operationally in communications. A prime example is \textit{%
teleportation} in which two different parties who share the state $\left|
\psi _{0}\right\rangle =\frac{1}{\sqrt{2}}\left( \left| 00\right\rangle
+\left| 11\right\rangle \right) $ are able to transfer an arbitrary quantum
state $\alpha \left| 0\right\rangle +\beta \left| 1\right\rangle $ from one
party to the other using classical communication and ``local'' operations.
Figuratively this means that $H^{\left( d_{1}\right) }$ with $d_{1}=2$ is
identified with one party, typically denoted as Alice, while $H^{\left(
d_{2}\right) }$ with $d_{2}=2$ is identified with a second party, typically
denoted as Bob. Alice and Bob can perform operations on their own components
of $H^{\left( N\right) }=H^{\left( d_{1}\right) }\otimes H^{\left(
d_{2}\right) }$ and can communicate classically which operations they
performed and whatever information they obtained from their operations.

An example of teleportation is the following. An arbitrary quantum state $%
\left| \psi \right\rangle =\alpha \left| 0\right\rangle +\beta \left|
1\right\rangle $ in a third Hilbert space is available to Alice. Without
actually knowing $\left| \psi \right\rangle $ she performs operations on
that state and on $H^{\left( d_{1}\right) }$ and, using classical
communication, transmits the results of a measurement to Bob who can then
recreate $\left| \psi \right\rangle $ in $H^{\left( d_{2}\right) }$, again
without knowing $\left| \psi \right\rangle $. (For references and
discussions see for example \cite{benn0,benn1,NC}.)

Two of the measures of entanglement for bipartite states which have been
motivated in part by teleportation are the \textit{measure of formation} and
the \textit{measure of distillation} (see for example \cite{benn1,benn2,woot}%
). The respective contexts concern the creation of a state from pure states
and ``distilling'' the maximum number of entangled states of the form $%
\left| \psi _{0}\right\rangle =\frac{1}{\sqrt{2}}\left( \left|
00\right\rangle +\left| 11\right\rangle \right) $ from copies of a given
density $\rho $. In a sense these particular measures can be considered as
operational measures, since they deal with the creation of a mixed state or
the extraction of maximally entangled pairs.

The definition of a measure of entanglement as an infimum of ``distance''
from $S$ was introduced in \cite{vedral2} with an expanded discussion in 
\cite{plenvid}. Other authors, particularly \cite{vidal1,vidal2}, took an
axiomatic approach and also discussed basic properties that such a measure
should possess, motivated in part by interpreting the operations in
teleportation and distillation as mappings of densities in $M$. (For an
alternate approach to entanglement using ``robustness of entanglement'' see 
\cite{vidal1}.)

As a paradigm based on \cite{vedral2}, we give the motivation (and
terminolgy) for requiring that a measure of entanglement $E$ satisfy the
following three properties: 
\begin{eqnarray*}
\text{(a) }E\left[ \rho \right] &=&0\text{ if and only if }\rho \in S\text{,}
\\
\text{(b) }E\left[ U\rho U^{\dagger }\right] &=&E\left[ \rho \right] \text{
where }U\text{ is a local unitary mapping,} \\
\text{(c) }E\left[ \Phi \left( \rho \right) \right] &\leq &E\left[ \rho
\right] \text{ where }\Phi \text{ is a local, completely positive,} \\
&&\text{ trace preserving operator on }D\text{.}
\end{eqnarray*}
The motivation for the first property is obvious if one is measuring
non-separability. A \textit{local} unitary mapping $U$ is a tensor product
of unitary maps on the constituent product spaces and models the unitary
transformations, such as a change of local basis, that could be taken
independently on the individual spaces. Property (b) requires that
entanglement should remain the same under such mappings. A completely
positive trace preserving operator $\Phi $ on $D$ models the measurement
process and can be represented \cite{Shu} as 
\[
\Phi \left( \rho \right) =\sum_{k}V_{k}\rho V_{k}^{\dagger }\text{ where }
\sum_{k}V_{k}^{\dagger }V_{k}=I. 
\]
It is easy to confirm that $\Phi $ maps $D$ into $D$. \textit{Locality} is
imposed by either assuming $V_{k}$ is a tensor product of operators on the
constituent spaces or else by simply assuming that $\Phi $ also maps $S$
into $S$. The point of axiom (c) is that one should not be able to increase
entanglement under local operations.

We mentioned above that the normalized identity $D_{0}$ is a separable
state, and it is obvious that $D_{0}$ can be written as an (equally
weighted) convex combination of any $N$ orthogonal projections, each of
which could be entangled. For example, in the two qubit case $D_{0}$ can be
written as the average of the densities defined by the four orthogonal
``Bell states'' $\frac{1}{\sqrt{2}}\left( \left| 00\right\rangle \pm \left|
11\right\rangle \right) $ and $\frac{1}{\sqrt{2}}\left( \left|
01\right\rangle \pm \left| 10\right\rangle \right) $. Thus one could expect
that a measure of entanglement would recognize the \textit{decrease} of
entanglement under convex combinations and satisfy 
\[
\text{(d) }E\left[ \sum p_{a}\rho _{a}\right] \leq \sum p_{a}E\left[ \rho
_{a}\right] \text{.} 
\]
This property is not a standard requirement, although it is just the
triangle inequality for distances, and it can be shown \cite{plenvid} that
some of the proposed measures automatically satisfy (d).

Since our goal here is to gain insight into the geometry of the separable
densities, we will not give a complete account of the various measures of
entanglement which have been proposed but instead will use the measure of
non-separability which comes naturally from the Hilbert space structure of $M
$. Invoking the idea of minimal distance from $S$ \cite{plenvid}, we use the 
\textit{Frobenius} or \textit{Hilbert-Schmidt} norm and define a measure of
entanglement as the minimal distance of a density $\rho $ from the set of
separable states: 
\begin{equation}
m\left[ \rho \right] =\inf_{\tau \in S}\left\| \rho -\tau \right\|
=\inf_{\tau \in S}\sqrt{Tr\left( \left( \rho -\tau \right) ^{2}\right) }.
\end{equation}
This measure has been already been considered as a possible measure of
entanglement by other authors such as \cite{steiner1,plenvid,witru}, but
since it does not seem to relate to operational uses of entanglement between
parties, it has not been widely used. However, it is easy to show that $m$
satisfies most of the properties discussed above. For example $m\left[ \rho
\right] =0$ if and only if $\rho $ is in $\mathit{S}$, and it is easy to
check that $m\left[ U\rho U^{-1}\right] =m\left[ \rho \right] $ for all
local unitary operations. $m$ also satisfies (d). Let $\rho =$ $%
\sum_{a}p_{a}\rho _{a}$ and suppose $m\left[ \rho _{a}\right] =\left\| \rho
_{a}-\tau _{a}\right\| $. Then by the definition and the triangle inequality 
\[
m\left[ \rho \right] \leq \left\| \sum p_{a}\left( \rho _{a}-\tau
_{a}\right) \right\| \leq \sum p_{a}\left\| \rho _{a}-\tau _{a}\right\|
=\sum p_{a}m\left[ \rho _{a}\right] .
\]

Although it does not seem to be known whether $m$ satisfies condition (c) as
stated, see \cite{ozawa,plenvid,witru} for discussions of this point, $m$
does satisfy a special case of (c) when $\Phi $ models a von Neumann
measurement. Specifically, we also assume that $\left\{ V_{k}\right\} $ is a
complete set of orthogonal projections which map $S$ to $S$ and let $\tilde{
\rho}=\Phi \left( \rho \right) $. If $\tau _{0}$ is the nearest separable
density to $\rho $, then 
\begin{eqnarray*}
\left\| \Phi \left( \rho \right) -\Phi \left( \tau _{0}\right) \right\| ^{2}
&=&Tr\left( \sum_{j}\sum_{k}V_{j}\left( \rho -\tau _{0}\right)
V_{j}^{\dagger }V_{k}\left( \rho -\tau _{0}\right) V_{k}^{\dagger }\right) \\
&=&Tr\left( \left( \sum_{j}V_{j}\left( \rho -\tau _{0}\right) V_{j}^{\dagger
}\right) \left( \rho -\tau _{0}\right) \right) \\
&\leq &\left\| \Phi \left( \rho \right) -\Phi \left( \tau _{0}\right)
\right\| \cdot \left\| \rho -\tau _{0}\right\| .
\end{eqnarray*}
Thus $m\left[ \tilde{\rho}\right] \leq \left\| \Phi \left( \rho \right)
-\Phi \left( \tau _{0}\right) \right\| \leq $ $\left\| \rho -\tau
_{0}\right\| =m\left[ \rho \right] $ as advertised. Since the inner product
structure of $M$ also gives geometric insights to aspects of the
separability problem, as shown for example in Witte and Truck's paper \cite
{witru}, we shall use $m$ as the measure of choice in this paper.

As a final remark on the issue of measures of entanglement, we note that
Vedral and Plenio \cite{plenvid} suggested that condition (c) be replaced by 
\[
\text{(c}^{\prime }\text{) }\sum_{k}p_{k}E\left[ \rho _{k}\right] \leq
E\left[ \rho \right] 
\]
where the $\rho _{k}$'s are particular densities derived from $\rho $ via $%
\Phi $: 
\[
\rho _{k}=\frac{1}{p_{k}}V_{k}\rho V_{k}^{\dagger }\text{ with }%
p_{k}=Tr\left( V_{k}\rho V_{k}^{\dagger }\right) . 
\]
They give a reasonable motivation for this stronger condition, but it should
be noted that if $\rho =\sum_{k}p_{k}\rho _{k}$ to begin with and if $\Phi $
is a von Neumann measurement leaving each $\rho _{k}$ unchanged, then if $E$
also satisfies the convexity property (d), $E$ has to be linear in this
case: 
\[
E\left[ \rho \right] =\sum_{k}p_{k}E\left[ \rho _{k}\right] \text{.} 
\]
A measure based on relative entropy does satisfy this condition \cite
{plenvid}, but it is too strong a condition for the Frobenius norm.

\section{\textbf{Basic Theory}}

In each of the $n$ Hilbert spaces $H^{\left( d_{k}\right) }$ defining $%
H^{\left( N\right) }$ we can define an orthogonal basis which arises from
the physical properties of the $d_{k}$-level system we are modelling. In the
quantum computation literature this is called the \textit{computational basis%
}, and tensor products of these basis vectors define a basis for $H^{\left(
N\right) }$. If we define projection operators on each of the $H^{\left(
d_{k}\right) }$, then their tensor products are the separable projections $%
\otimes _{k}\left| \psi _{k}\right\rangle \left\langle \psi _{k}\right| $
whose convex hull is $\mathit{S}$. More generally, if we take a basis for
the $d_{k}^{2}$ dimensional space of linear operators on $H^{\left(
d_{k}\right) }$, then their tensor products define a basis for the $N^{2}$
dimensional space $M$.

Now it was shown in \cite{pitrub2} that one can take what amounts to a
discrete Fourier transform of a suitable arrangement of such product basis
matrices and obtain a particular orthogonal unitary basis $\left\{ S_{\left(
j,k\right) }\right\} $ for $\mathit{M}$ which is indexed by pairs $\left(
j,k\right) $ of $n$-long vectors $j,k$. Using coordinate-wise addition, the
set of indices defines an Abelian group $G$ of order $N$, and $\left\{
S_{\left( j,k\right) }\right\} $ turns out to be a projective representation
of $G\times G$. The $S_{\left( j,k\right) }$ are unitary matrices and need
not be in $\mathit{M}$; rather they serve as a basis of $N\times N$ matrices
over the complex numbers. The reader is referred to \cite{pitrub2} for
details of the construction, and we limit ourselves here to recording the
results we need. (See \cite{fivel,knill}. We also note that Werner \cite
{werner2} shows the close connections among orthogonal unitary bases, dense
coding and teleportation, all topics of great interest in quantum computing.)

Using $e$ to denote $\left( 0,0\right) $ and $a$ as a generic index $\left(
j,k\right) $, the unitary matrices in $\left\{ S_{a}\right\} $ have the
following properties: (1) $S_{e}=I$, the $N\times N$ identity, (2) $Tr\left(
S_{a}^{\dagger }S_{b}\right) =N\delta \left( a,b\right) $, and (3) $S_{a}$
has the spectral representation $\sum_{k}\lambda _{a,k}P_{a,k},$ where the $%
P_{a,k}$ are \textit{separable} orthogonal projections. Since $S_{a}$ is
unitary, $\sum_{k}P_{a,k}=I$ and $\left| \lambda _{a,k}\right| =1$. Since $%
\left\{ S_{a}\right\} $ is a basis, a density $\rho $ in $\mathit{D}$ can be
expressed as $\frac{1}{N}\sum_{a}s_{a}S_{a}$, and the last particular
property is that $\left| s_{a}\right| \leq 1$ and $s_{e}=1$.

It has been shown in a number of papers, initially in \cite{zycz} and also
in \cite{braun,steiner1,pitrub2} for example, that there is an open
neighborhood of the normalized identity $D_{0}$ which is composed entirely
of separable densities. Using the properties of $\left\{ S_{a}\right\} $ we
give a short proof.

\begin{proposition}
If $\rho $ is a density with $\sum_{a\neq e}\left| s_{a}\right| \leq 1$,
then $\rho $ is separable. In particular, there exists an open neighborhood
of $D_{0}$ composed of separable states.
\end{proposition}

\textit{proof}: Since $\rho $ is Hermitian, $\rho =\frac{1}{2}\left( \rho
+\rho ^{\dagger }\right) $. Using the various properties listed above we
have 
\begin{eqnarray*}
\rho &=&\frac{1}{N}\left[ S_{e}+\frac{1}{2}\sum_{a\neq e}\left( s_{a}S_{a}+%
\bar{s}_{a}S_{a}^{\dagger }\right) \right] \\
&=&\frac{1}{N}\left[ S_{e}+\sum_{a\neq e}\sum_{k}\frac{1}{2}\left(
s_{a}\lambda _{a,k}+\bar{s}_{a}\bar{\lambda}_{a,k}\right) P_{a,k}\right] \\
&=&\left( 1-\sum_{a\neq e}\left| s_{a}\right| \right) \frac{1}{N}%
S_{e}+\sum_{a\neq e}\sum_{k}\left| s_{a}\right| \frac{1}{N}\left( 1+\cos
\left( \theta _{a,k}\right) \right) P_{a,k}\text{,}
\end{eqnarray*}
where $\frac{1}{2}\left( s_{a}\lambda _{a,k}+\bar{s}_{a}\bar{\lambda}%
_{a,k}\right) =$ $\left| s_{a}\right| \cos \left( \theta _{a,k}\right) $.
Since $1+\cos \left( \theta _{a,k}\right) \geq 0$, we have written $\rho $
explicitly as a convex combination of separable densities, and thus $\rho $
is in $\mathit{S}$. For the last assertion the condition $\sum_{a\neq
e}\left| s_{a}\right| <1$ defines a relatively open set in $D$. 
$\square $

In order to determine if an individual density is separable using this
criterion, one has to compute each of the coefficients $s_{a}$. Two weaker
but user friendly corollaries are immediate consequences, however.

\begin{corollary}
If $\epsilon <\left( N^{2}-1\right) ^{-1}$, then $\left\{ \left( 1-\epsilon
\right) D_{0}+\epsilon \sigma ,\sigma \in \mathit{D}\right\} $ is a
relatively open set of densities in $\mathit{S}$.
\end{corollary}

\textit{Proof}: Let $\mu $ denote a density $\left( 1-\epsilon \right)
D_{0}+\epsilon \sigma $, so that the $a\neq e$ coefficient of $\mu $ is $%
\epsilon s_{a}$ where $s_{a}$ is the corresponding coefficient of $\sigma $.
Since $\epsilon <\left( N^{2}-1\right) ^{-1}$, $\sum_{a\neq e}\epsilon
\left| s_{a}\right| <1$ and $\mu $ is separable.\ $\square $

Generally speaking it does not appear that the eigenvalues and eigenvectors
of a density are useful in distinguishing a separable from a non-separable
state. One counterexample is the following result which is not particularly
strong but which has an easy proof.

\begin{corollary}
If the smallest eigenvalue of a density $\rho $ is at least $\frac{1}{N+t}$
where $t=\frac{N}{N^{2}-2}$, then $\rho $ is separable.
\end{corollary}

\textit{Proof}: We can use the spectral representation of $\rho $, obtaining 
\begin{eqnarray*}
\rho &=&\sum_{k}\left( \lambda _{k}-\frac{1}{N+t}\right) \left| \psi
_{k}\right\rangle \left\langle \psi _{k}\right| +\frac{N}{N+t}\sum_{k}\frac{1%
}{N}\left| \psi _{k}\right\rangle \left\langle \psi _{k}\right| \\
&=&\frac{t}{N+t}\sum_{k}\alpha _{k}\left| \psi _{k}\right\rangle
\left\langle \psi _{k}\right| +\frac{N}{N+t}D_{0},
\end{eqnarray*}
where $0\leq \alpha _{k}=\frac{\lambda _{k}\left( N+t\right) -1}{t}$. Thus $%
\mu =\sum_{k}\alpha _{k}\left| \psi _{k}\right\rangle \left\langle \psi
_{k}\right| $ is a density. In the unitary basis 
\[
\rho =\frac{1}{N}\left[ S_{e}+\frac{t}{N+t}\sum_{a\neq e}s_{a}S_{a}\right] 
\]
where the $s_{a}$ are the $\left\{ S_{a}\right\} $ coefficients of $\rho $.
Then 
\[
\frac{t}{N+t}\sum_{b\neq e}\left| s_{b}\right| \leq \frac{t}{N+t}\left(
N^{2}-1\right) \leq 1 
\]
showing that $\rho $ is separable.\ $\square $

As an example of Proposition 4.1, one can use the definition of the $S_{a}$%
's as in \cite{pitrub1} to show that the $s_{a}$ coefficients of the two
qubit density $\rho \left( s\right) $ defined above satisfy $\sum_{a\neq
e}\left| s_{a}\right| =3s$. Thus $s\leq \frac{1}{3}$ is also a sufficient
condition for separability, and one does not need the Horodecki-Peres
result. Our main application of the preceding proposition, however, is to
characterize densities in the relative interiors of $\mathit{S}$ and $%
\mathit{D}$.

\begin{proposition}
A density $\rho $ in $\mathit{S}$ is in the relative interior of $\mathit{S}$
if and only if there exists a $t>0$ such that $\left( 1+t\right) \rho
-tD_{0} $ is in $S$. The same assertion holds if $\mathit{S}$ is replaced by 
$\mathit{D}$ throughout.
\end{proposition}

\textit{Proof}: Suppose that $\mu =\left( 1+t\right) \rho -tD_{0}$ is in $%
\mathit{S}$. Then for any $\sigma $ in $\mathit{D}$%
\[
\frac{1}{1+t}\mu +\frac{t}{1+t}\left( D_{0}+\epsilon \left( \sigma
-D_{0}\right) \right) =\rho +\frac{t\epsilon }{1+t}\left( \sigma
-D_{0}\right) 
\]
is also in $\mathit{S}$ provided $\epsilon <\left( N^{2}-1\right) ^{-1}$.
Conversely, if $\rho $ is in the relative interior of $\mathit{S}$, then for
small $\delta $, $\rho +\delta \left( \sigma -D_{0}\right) $ is in $\mathit{S%
}$ for all $\sigma $ in $\mathit{D}$ so that choosing $\sigma =\rho $ gives
a separable density $\mu =\left( 1+\delta \right) \rho -\delta D_{0}$. The
same proof works if $\mathit{S}$ is replaced by $\mathit{D}$.\ $\square $

The use of a line segment connecting a density with the normalized identity $%
D_{0}$ turns out to be a helpful tool in the analysis. Accordingly we shall
refer to $\left( 1+t\right) \rho -tD_{0}$ as an \textit{entanglement probe}
and note that Vidal and Tarrach \cite{vidal1} made extensive use of
entanglement probes in defining and investigating a ``robustness'' of
entanglement for densities. Two easy applications show both the utility of
entanglement probes and the contrast between $\mathit{S}$ and $\mathit{D}$.

\begin{corollary}
A density $\rho $ is on the boundary of $\mathit{D}$ if and only if $\rho $
has a zero eigenvalue. If $\rho $ is in $\mathit{S}$ and rank$\left( \rho
\right) <N$, then $\rho $ is on the boundary of $\mathit{S}$ and also of $%
\mathit{D}$.
\end{corollary}

\textit{Proof}: If $\rho \left| \psi \right\rangle =0\left| \psi
\right\rangle $, then $\left( 1+t\right) \left\langle \psi \left| \rho
\right| \psi \right\rangle -t\left\langle \psi \left| D_{0}\right| \psi
\right\rangle <0$ and $\mu =\left( 1+t\right) \rho -tD_{0}$ is not positive
semidefinite. Conversely, if the eigenvalues of $\rho $ are bounded below by 
$s>0,$ then $\rho $ can be written as 
\[
\rho =\left( 1-sN\right) \sum_{k}\frac{\lambda _{k}-s}{1-sN}\left| \psi
_{k}\right\rangle \left\langle \psi _{k}\right| +sND_{0} 
\]
and since $sN\leq 1$, $\rho $ is in the interior of $D$. If $\rho $ is in $%
\mathit{S}$ and rank$\left( \rho \right) <N$, the same proof shows that $%
\rho $ is on the boundary of $\mathit{D}$ and thus also of $\mathit{S}$.\ $%
\square $

There is a class of densities which satisfy the Peres partial transform
condition but which are not separable, and there have been a number of
detailed investigations of these densities using rather different techniques
than those described above. For an introduction and additional references
see Lewenstein et al in \cite{Lew2}.

\section{\textbf{Orthogonality}}

In our running example we have seen that $\rho \left( \frac{1}{3}\right) $
on $H^{\left[ 2\right] }\otimes H^{\left[ 2\right] }$ is the closest
separable density to the inseparable density $\rho _{0}$ along the line
connecting $D_{0}$ and $\rho _{0}$. As it happens, $\rho \left( \frac{1}{3}%
\right) $ is also closest to $\rho _{0}$ in the norm $\left\| \;\right\| $
defined in equation (3.1). To see this we need an alternate characterization
of the density in $\mathit{S}$ closest to a density $\rho $ in $\mathit{D}-%
\mathit{S}$. This characterization is a standard result in convexity theory
and has been used in \cite{steiner1,steiner2,witru} for example.

\begin{proposition}
Suppose $\rho $ in $\mathit{D}-\mathit{S}$. Then $\tau _{0}$ is the unique
closest separable density to $\rho $ if and only if for all $\tau $ in $%
\mathit{S}$%
\begin{equation}
\left\langle \left\langle \rho -\tau _{0},\tau -\tau _{0}\right\rangle
\right\rangle \equiv Tr\left( \left( \rho -\tau _{0}\right) \left( \tau
-\tau _{0}\right) \right) \leq 0.
\end{equation}
By the convexity of $S$, it suffices to prove the inequality for all
separable projections $\tau $.
\end{proposition}

\textit{Proof}: Adding and subtracting $\tau _{0}$ gives 
\[
\left\langle \left\langle \rho -\tau ,\rho -\tau \right\rangle \right\rangle
=\left\langle \left\langle \rho -\tau _{0},\rho -\tau _{0}\right\rangle
\right\rangle -2\left\langle \left\langle \rho -\tau _{0},\tau -\tau
_{0}\right\rangle \right\rangle +\left\langle \left\langle \tau _{0}-\tau
,\tau _{0}-\tau \right\rangle \right\rangle 
\]
which shows (5.1) is sufficient. Conversely, if $\left\langle \left\langle
\rho -\sigma ,\rho -\sigma \right\rangle \right\rangle $ is minimal over $%
\mathit{S}$ when $\sigma =\tau _{0}$, then $\left\langle \left\langle \rho
-\tau _{0},\sigma -\tau _{0}\right\rangle \right\rangle \leq \frac{1}{2}
\left\langle \left\langle \tau _{0}-\sigma ,\tau _{0}-\sigma \right\rangle
\right\rangle $. Using the convexity of the separable states, let $\sigma
=\left( 1-t\right) \tau +t\tau _{0}$ with $0<t<1$, where $\tau $ is in $%
\mathit{S}$. It follows that 
\[
\left\langle \left\langle \rho -\tau _{0},\tau -\tau _{0}\right\rangle
\right\rangle \leq \frac{1}{2}\left( 1-t\right) \left\langle \left\langle
\tau _{0}-\tau ,\tau _{0}-\tau \right\rangle \right\rangle \text{,} 
\]
and letting $t$ go to one gives the result. If $\tau _{1}$ also minimizes $%
\left\langle \left\langle \rho -\tau ,\rho -\tau \right\rangle \right\rangle 
$, then from $\left\langle \left\langle \rho -\tau _{0},\tau _{1}-\tau
_{0}\right\rangle \right\rangle \leq 0$ and $\left\langle \left\langle \rho
-\tau _{1},\tau _{0}-\tau _{1}\right\rangle \right\rangle \leq 0$, we can
conclude that $\left\langle \left\langle \tau _{1}-\tau _{0},\tau _{1}-\tau
_{0}\right\rangle \right\rangle \leq 0$, confirming uniqueness and
completing the proof.\ $\square $

An extremely useful geometric entity is the separable face nearest a given $%
\rho _{0}$ in $\mathit{D}-\mathit{S}$. Let $\tau _{0}$ denote the nearest
separable density to $\rho _{0}$ and use the notation of the proposition
above.

\begin{definition}
$F\left( \rho _{0},\tau _{0}\right) $ denotes $\left\{ \tau \in \mathit{S}%
:\left\langle \rho _{0}-\tau _{0},\tau -\tau _{0}\right\rangle =0\right\} .$
\end{definition}

Thus $F\left( \rho _{0},\tau _{0}\right) $ is the convex set of densities in 
$\mathit{S}$ such that as vectors $\tau -\tau _{0}$ is perpendicular to $%
\rho _{0}-\tau _{0}$. We leave it to the reader to confirm that $F\left(
\rho _{0},\tau _{0}\right) $ is indeed a face of $\mathit{S}$ and that the
extreme separable projections in a convex representation of $\tau _{0}$
necessarily lie in $F\left( \rho _{0},\tau _{0}\right) $.

The alternate characterization of $\tau _{0}$ allows us to compute the
nearest separable density in some cases, and we pursue that idea next. As an
example, the following result includes Proposition 1 of \cite{witru} as a
special case in which the density $\rho _{1}$ below is separable and equal
to a density in $F=F\left( \rho _{0},\tau _{0}\right) $.

\begin{corollary}
Suppose $\tau _{0}$ and $\tau _{1}$, the nearest separable densities to $%
\rho _{0}$ and $\rho _{1}$ respectively, are both in $F=F\left( \rho
_{0},\tau _{0}\right) $. Then the nearest separable density to $t\rho
_{0}+\left( 1-t\right) \rho _{1}$ is $\tau \left( t\right) =t\tau
_{0}+\left( 1-t\right) \tau _{1}$, and thus $m\left[ t\rho _{0}+\left(
1-t\right) \rho _{1}\right] $ $\leq tm\left[ \rho _{0}\right] +\left(
1-t\right) m\left[ \rho _{1}\right] $.
\end{corollary}

\textit{Proof}: Since $\left\langle \left\langle \rho _{0}-\tau _{0},\tau
\left( t\right) -\tau _{0}\right\rangle \right\rangle =\left\langle
\left\langle \rho -\tau _{1},\tau \left( t\right) -\tau _{1}\right\rangle
\right\rangle =0$, we have 
\begin{eqnarray*}
&&\left\langle \left\langle t\rho _{0}+\left( 1-t\right) \rho _{1}-\tau
\left( t\right) ,\tau -\tau \left( t\right) \right\rangle \right\rangle \\
&=&t\left\langle \left\langle \rho _{0}-\tau _{0},\tau -\tau \left( t\right)
\right\rangle \right\rangle +\left( 1-t\right) \left\langle \left\langle
\rho _{1}-\tau _{1},\tau -\tau \left( t\right) \right\rangle \right\rangle \\
&=&t\left\langle \left\langle \rho _{0}-\tau _{0},\tau -\tau
_{0}\right\rangle \right\rangle +\left( 1-t\right) \left\langle \left\langle
\rho _{1}-\tau _{1},\tau -\tau _{1}\right\rangle \right\rangle \leq 0\text{,}
\end{eqnarray*}
completing the proof.\ $\square $

As another application, we are able to give a geometric perspective to $\tau
_{0}\left( d\right) $, the separable density closest to the bipartite state $%
\rho _{0}\left( d\right) =\left| \psi _{d}\right\rangle \left\langle \psi
_{d}\right| $ where 
\[
\left| \psi _{d}\right\rangle =\left| \psi _{d}\left( 2\right) \right\rangle
=\frac{1}{\sqrt{d}}\sum_{j=0}^{d-1}\left| jj\right\rangle .
\]
This includes the motivating example as a special case. The state $\rho
_{0}\left( d\right) $ is known as a maximally entangled state and for $d=2$
was used by Werner \cite{werner1} in an analysis of ``local reality'' and
the Einstein, Podolsky, Rosen paradox \cite{epr}. Now the convex combination 
$\left( 1-s\right) D_{0}+s\rho _{0}\left( d\right) $ can be interpreted as a
mixture of the maximally entangled state $\rho _{0}\left( d\right) $ and $%
D_{0}$, the ``maximally mixed'' state or random noise. In earlier studies,
including \cite{pitrub3} and \cite{rungta} and references therein, the
largest value of $s$ for which $\left( 1-s\right) D_{0}+s\rho _{0}\left(
d\right) $ is separable was investigated, which is equivalent to the
question of how much noise it takes to make the system unentangled. As it
happens, the nearest separable density to $\rho _{0}\left( d\right) $ is
such a convex combination. That result for $d=2$ seems to have been noticed
first in \cite{witru}, and the proof below for arbitrary $d$ follows their
approach. (An independent proof of the general case recently appeared as
part of the analysis in \cite{steiner1}.)

\begin{proposition}
The state $\tau _{0}\left( d\right) =\left( 1-s_{d}\right) D_{0}+s_{d}\rho
_{0}\left( d\right) $ with $s_{d}=\left( 1+d\right) ^{-1}$ is the nearest
separable density to the maximally entangled state $\rho _{0}\left( d\right) 
$. The analogous assertion is false if the number of product states $n$ is
bigger than $2$.
\end{proposition}

\textit{Proof}: The proof that $\tau _{0}\left( d\right) $ is separable has
been given in a number of references such as \cite{braun,rungta,pitrub3,duer}
among others. Dropping explicit mention of $d$, we thus need to prove that
for all separable projections $\tau $ 
\[
\frac{1}{\left( 1-s_{d}\right) }\left\langle \left\langle \rho _{0}-\tau
_{0},\tau -\tau _{0}\right\rangle \right\rangle =\left\langle \left\langle
\rho _{0}-D_{0},\tau -\tau _{0}\right\rangle \right\rangle =\left\langle
\left\langle \rho _{0},\tau -\tau _{0}\right\rangle \right\rangle \leq 0. 
\]
First, $Tr\left( \left| \psi _{d}\right\rangle \left\langle \psi _{d}\right|
\tau _{0}\right) =\left( 1-s_{d}\right) Tr\left( \frac{1}{N}\left| \psi
_{d}\right\rangle \left\langle \psi _{d}\right| \right) +s_{d}Tr\left(
\left| \psi _{d}\right\rangle \left\langle \psi _{d}\right| \right) =\frac{d%
}{1+d}\frac{1}{d^{2}}+\frac{1}{1+d}=\frac{1}{d}$. If $\tau $ is a separable
projection, then $\tau =\left| \alpha \right\rangle \left\langle \alpha
\right| \otimes \left| \beta \right\rangle \left\langle \beta \right| $
where in the computational basis $\left| \alpha \right\rangle
=\sum_{k=0}^{d-1}a_{k}\left| k\right\rangle $ with $\sum_{k}\left|
a_{k}\right| ^{2}=1$ and with an analogous expression for $\left| \beta
\right\rangle $. Then 
\[
Tr\left( \left| \psi _{d}\right\rangle \left\langle \psi _{d}\right| \tau
\right) =\frac{1}{d}\left| \sum_{i,j,k}\left\langle ii|jk\right\rangle
a_{j}b_{k}\right| ^{2}=\frac{1}{d}\left| \sum_{k}a_{k}b_{k}\right| ^{2}\leq 
\frac{1}{d} 
\]
since $\left| \sum_{k}a_{k}b_{k}\right| ^{2}\leq \sum_{k}\left| a_{k}\right|
^{2}\sum_{k}\left| \bar{b}_{k}\right| ^{2}$ by the Cauchy-Schwarz
inequality. Hence $\left\langle \left\langle \rho _{0}-\tau _{0},\tau -\tau
_{0}\right\rangle \right\rangle \leq 0$ for all separable densities and $%
\tau _{0}\left( d\right) $ is the closest separable density to $\rho
_{0}\left( d\right) $ when $n=2$. When $n>2$ and $\left| \psi _{d}\left(
n\right) \right\rangle $ is defined analogously, $Tr\left( \left| \psi
_{d}\left( n\right) \right\rangle \left\langle \psi _{d}\left( n\right)
\right| \tau _{0}\right) $ equals $\frac{1+d}{d\left( 1+d^{n-1}\right) }$
and it is easy to see that there are separable projections with $Tr\left(
\left| \psi _{d}\left( n\right) \right\rangle \left\langle \psi _{d}\left(
n\right) \right| \tau \right) =\frac{1}{d}$, completing the proof of the
proposition.\ $\square $

\begin{corollary}
Using $m$ as the measure of entanglement, $m\left[ \rho _{0}\left( d\right)
\right] =\sqrt{1-\frac{2}{d+1}}$, so that entanglement increases with
increasing $d$.\ $\square $
\end{corollary}

As another application, we can compute explicitly the extreme points of $%
F\left( \rho _{0}\left( d\right) ,\tau _{0}\left( d\right) \right) $.

\begin{corollary}
$F\left( \rho _{0}\left( d\right) ,\tau _{0}\left( d\right) \right) $ is the
convex hull of $\left| \alpha \right\rangle \left\langle \alpha \right|
\otimes \left| \bar{\alpha}\right\rangle \left\langle \bar{\alpha}\right| $,
where the bar denotes the complex conjugate of the entries of the row or
column vector.
\end{corollary}

\textit{Proof}: From an earlier observation, it suffices to consider
densities of the form $\tau =\left| \alpha \right\rangle \left\langle \alpha
\right| \otimes \left| \beta \right\rangle \left\langle \beta \right| $.
From the proof above, $\tau $ is in $F\left( \rho _{0}\left( d\right) ,\tau
_{0}\left( d\right) \right) $ if and only if $\left|
\sum_{k}a_{k}b_{k}\right| ^{2}=\sum_{k}\left| a_{k}\right|
^{2}\sum_{k}\left| \bar{b}_{k}\right| ^{2}$. This is the case of equality in
the Cauchy-Schwarz inequality over the complex numbers and is equivalent to $%
b_{k}=c\bar{a}_{k}$ for some constant $c$ and all $k$. (See for example \cite
{rudin}.) By the normalization condition, $\left| c\right| =1$ and thus does
not appear as a factor in $\tau =\left| \alpha \right\rangle \left\langle
\alpha \right| \otimes \left| \bar{\alpha}\right\rangle \left\langle \bar{
\alpha}\right| $, completing the proof.\ $\square $

As an example, when $d=n=2$, the basis of orthogonal unitary matrices $%
\left\{ S_{a}\right\} $ defined earlier is essentially the set of four Pauli
matrices: $\sigma _{0}=\left( 
\begin{array}{ll}
1 & 0 \\ 
0 & 1
\end{array}
\right) $, $\sigma _{z}=\left( 
\begin{array}{cc}
1 & 0 \\ 
0 & -1
\end{array}
\right) $, $\sigma _{x}=\left( 
\begin{array}{ll}
0 & 1 \\ 
1 & 0
\end{array}
\right) $, and $\sigma _{y}=\left( 
\begin{array}{cc}
0 & -i \\ 
i & 0
\end{array}
\right) $. (The only difference is that one uses $\left( 
\begin{array}{cc}
0 & 1 \\ 
-1 & 0
\end{array}
\right) $ in lieu of $\sigma _{y}$.) It is easy to show that $\tau _{0}=\rho
\left( \frac{1}{3}\right) $ is the average of the six separable projections $%
\frac{1}{4}\left( \sigma _{0}\pm \sigma _{z}\right) \otimes \left( \sigma
_{0}\pm \sigma _{z}\right) $, $\frac{1}{4}\left( \sigma _{0}\pm \sigma
_{x}\right) \otimes \left( \sigma _{0}\pm \sigma _{x}\right) $, and $\frac{1%
}{4}\left( \sigma _{0}\pm \sigma _{y}\right) \otimes \left( \sigma _{0}\mp
\sigma _{y}\right) $ and that these projections have the requisite form $%
\left| \alpha \right\rangle \left\langle \alpha \right| \otimes \left| \bar{
\alpha}\right\rangle \left\langle \bar{\alpha}\right| $.

\section{\textbf{Entanglement witnesses}}

Suppose $\rho _{0}\notin S$. Then a standard consequence of the Hahn-Banach
theorem for convex spaces is that there exists a linear functional $F$ on $M$
such that $F\left( \rho _{0}\right) <0\leq \inf \left[ F\left( \tau \right)
,\tau \in S\right] $. In the context of our finite dimensional Hilbert space 
$M$, the Riesz representation theorem says that each linear functional is of
the form $F\left( \rho \right) =Tr\left( A\rho \right) $ for some Hermitian
matrix $A.$ (See for example \cite{Ax}.) It is also a standard fact that the
hyperplane $M\left( A\right) =\left\{ B\in M:Tr\left( AB\right) =0\right\} $
has dimension $\dim \left( M\right) -1$, so that one can view $M\left(
A_{0}\right) $ as a separating hyperplane with $\rho _{0}$ on one side and $%
S $ on the other. Since quantum mechanical \textit{observables }are modelled
as Hermitian matrices, the thrust of the theory is that the condition $\rho
\notin S$ can be ``witnessed'' by a suitable observable $A$, and such
Hermitian matrices have been dubbed ``entanglement witnesses'' in the
quantum computation literature$.$ This connection was first pointed out in 
\cite{horo1}, and the authors went on to link these ideas to the Banach
algebra literature. In particular they showed that the Peres necessary
condition for separability is also sufficient in the $2\times 2$ and $%
2\times 3$ tensor product cases.

The Peres condition can be couched in the language of positive operators on
bounded functions on $H^{\left[ N\right] }$, and, as mentioned in the
introduction, one direction of research on separability has focused on
densities which satisfy the Peres condition but which are not separable. A
consequence of that work has been a study of entanglement witnesses in
general. Recent relevant papers which contain further references include 
\cite{Lew1,Lew2,Lew3,Ter1,Ter2}.

Since separating hyperplanes are not unique, it is customary to normalize in
the entanglement context by requiring that $Tr\left( AD_{0}\right) =1$ in
addition to 
\begin{equation}
Tr\left( A\rho _{0}\right) <0\leq \inf \left[ Tr\left( A\tau \right) ,\tau
\in S\right]
\end{equation}
for some inseparable density $\rho _{0}$. In \cite{Lew3} the authors
introduced a partial order on such entanglement witnesses as follows. Let $%
D\left( A\right) $ denote $\left\{ \rho \in D:Tr\left( A\rho \right)
<0\right\} $. Define a partial order by $A\preceq B$ if and only if $D\left(
A\right) \subseteq D\left( B\right) $. Then an \textit{optimal} entanglement
witness is a maximal element in the partial order. We should note that the
analysis in \cite{Lew3} deals with general entanglement witnesses, and one
of the sufficient conditions below for $A_{0}$ to be optimal appears there.

The connection with our analysis is that knowing the closest separable
density $\tau _{0}$ to a nonseparable density $\rho _{0}$ also enables one
to construct an entanglement witness $A_{0}$ for a class of densities
related to $\rho _{0}$. This is not an assertion that actually finding $\tau
_{0}$ is computationally easy. Rather, it illustrates the importance of $%
\tau _{0}$ and shows $Tr\left( A_{0}\sigma \right) $ has a familiar form
which further reveals its geometric character.

We assume equation (6.1), but since we begin with a particular $\rho _{0}$
we use a slightly different normalization.

\begin{definition}
$A_{0}$ is said to be \textit{optimal} provided that any Hermitian $A$
satisfying equation (6.1) together with $Tr\left( A\rho _{0}\right) =$ $%
Tr\left( A_{0}\rho _{0}\right) $ and $D\left( A_{0}\right) \subseteq D\left(
A\right) $ necessarily equals $A_{0}$.
\end{definition}

\begin{theorem}
Suppose $\tau _{0}$ is the nearest separable density to a non-separable
density $\rho _{0}$. Then the Hermitian matrix $A_{0}=c_{0}I+\tau _{0}-\rho
_{0}$ with $c_{0}=Tr\left( \tau _{0}\left( \rho _{0}-\tau _{0}\right)
\right) $ is an entanglement witness for $\rho _{0}$. In particular for any
density $\sigma $ 
\begin{equation}
Tr\left( A_{0}\sigma \right) =-\left\langle \left\langle \rho _{0}-\tau
_{0},\sigma -\tau _{0}\right\rangle \right\rangle 
\end{equation}
so that the separating hyperplane defined by $A_{0}$ contains $F\left( \rho
_{0},\tau _{0}\right) $. If some $\tau $ in $F\left( \rho _{0},\tau
_{0}\right) $ has full rank, then $A_{0}$ is optimal.
\end{theorem}

\textit{Proof}: It is easy to check equation (6.2) so that $A_{0}$ has the
asserted properties. Next, suppose that $A$ satisfies equation (6.1) and
that $D\left( A_{0}\right) \subseteq D\left( A\right) $ with $Tr\left( A\rho
_{0}\right) =$ $Tr\left( A_{0}\rho _{0}\right) $. Using one of the
techniques in \cite{Lew3}, suppose that $Tr\left( A_{0}\rho \right) =0$.
Then for $0<s<1$, $Tr\left( A_{0}\left( \left( 1-s\right) \rho +s\rho
_{0}\right) \right) <0$ implying $Tr\left( A\rho \right) <\frac{-s}{1-s}%
Tr\left( A\rho _{0}\right) $ and thus $Tr\left( A\rho \right) \leq 0$. In
particular $Tr\left( A\tau \right) \leq 0$ for $\tau $ in $F\left( \rho
_{0},\tau _{0}\right) $, forcing $Tr\left( A\tau \right) =0$.

Now suppose that there is a $\tau $ in $F\left( \rho _{0},\tau _{0}\right) $
with full rank, so that its smallest eigenvalue is strictly positive. Then
it's straightforward to show that there exists a small positive $t$ such
that for any density $\rho $, $\mu \left( t\right) =\left( 1+t\right) \tau
-t\rho $, a variant of the entanglement probes defined earlier, is in $D$.
In particular if $Tr\left( A_{0}\rho \right) =0$ we have 
\[
Tr\left( A_{0}\mu \left( t\right) \right) =0\geq Tr\left( A\mu \left(
t\right) \right) =-tTr\left( A\rho \right) \geq 0 
\]
forcing $Tr\left( A\rho \right) =0$. This gives the property that $Tr\left(
A_{0}\rho \right) =0$ implies $Tr\left( A\rho \right) =0$ which suffices for
the rest of the proof. In fact, that property together with equation (6.1)
and $Tr\left( A\rho _{0}\right) =$ $Tr\left( A_{0}\rho _{0}\right) $ is
equivalent to $A=A_{0}$.

Suppose $Tr\left( A_{0}\rho \right) >0.$ Then $Tr\left( A_{0}\left( \left(
1-s\right) \rho +s\rho _{0}\right) \right) =0$ for some $s$ in $\left(
0,1\right) $, and from the normalization it follows that $Tr\left( A_{0}\rho
\right) =Tr\left( A\rho \right) $. In particular $Tr\left( A_{0}D_{0}\right)
=Tr\left( AD_{0}\right) $. Finally, if $Tr\left( A_{0}\rho \right) <0$, then
analogously $Tr\left( A_{0}\left( \left( 1-s\right) \rho +sD_{0}\right)
\right) =0$ for some $s$ in $\left( 0,1\right) $, and that gives $Tr\left(
A_{0}\rho \right) =Tr\left( A\rho \right) $. Consequently $Tr\left(
A_{0}\rho \right) =Tr\left( A\rho \right) $ for all $\rho $ in $D$, and it
follows that $A_{0}=A$, completing the proof. \ $\square $

\begin{corollary}
$A_{0}$ is optimal if the separable eigenvectors of the rank one separable
projections in $F\left( \rho _{0},\tau _{0}\right) $ span $H^{\left(
N\right) }$ or if there is a density $\rho $ of full rank such that $\rho
=(1+t)\tau -t\rho _{0}$ for some $\tau $ in $F\left( \rho _{0},\tau
_{0}\right) $.
\end{corollary}

\textit{Proof}: In the first case, it is easy to see that one can construct
a $\tau $ in $F\left( \rho _{0},\tau _{0}\right) $ which has full rank. In
the second case, the techniques in Corollary (4.3) show that $\tau $ has
full rank.\ $\square $

As an example of this general theory we have the following specific result
which includes example 5, up to a multiplicative constant, in \cite{Ter2}.

\begin{corollary}
For the usual $2\times 2$ bivariate example, $\tau _{0}=\rho \left(
1/3 \right) $ has full rank and 
\[
A_{0}=\frac{1}{3}\left( I-2\rho _{0}\right) =\frac{1}{3}\left( 
\begin{array}{cccc}
0 & 0 & 0 & -1 \\ 
0 & 1 & 0 & 0 \\ 
0 & 0 & 1 & 0 \\ 
-1 & 0 & 0 & 0
\end{array}
\right) 
\]
is an optimal entanglement witness. $D\left( A_{0}\right) $ contains $\rho
_{a}=\left| \psi _{a}\right\rangle \left\langle \psi _{a}\right| $ for any
density of the form $\rho _{a}=\left| \psi _{a}\right\rangle \left\langle
\psi _{a}\right| $ where $\left| \psi _{a}\right\rangle
=\sum_{k=0}^{1}a_{k}\left| kk\right\rangle $ with non-negative $a_{k}$ such
that $\sum_{k}\left| a_{k}\right| ^{2}=1$.\ In the corresponding $d\times d$
case, $A_{0}=\frac{1}{1+d}\left( I-d\rho _{0}\right) $. $\square $
\end{corollary}

If the matrix above is denoted as $M_{01}$, the $A_{0}$ in the $d\times d$
case turns out to be a multiple of $\sum_{0\leq j<k<d}M_{jk}$, where the $%
M_{jk}$ have definitions analogous to $M_{01}$. (See equation (7.4).)
Another role for the $M_{jk}$ is given below, where we find other nearest
separable states using an extension of the methodology developed above.

\section{\textbf{Variations in the bivariate case.}}

It would be useful to be able to calculate $m\left[ \rho \right] $, the
Frobenius measure of entanglement of states other than the maximally
entangled states, and we can do this for states which are near to the
maximally entangled state in a sense to be made more precise below. We will
use the geometric insights obtained above in the context of two $d$-level
systems and motivate the analysis with the usual two qubit case $d=2$. That
particular case was also studied by Witte and Trucks \cite{witru} who used a
different approach to obtain Proposition 7.1 below.

Our approach is motivated by the geometry. We know that $\rho _{0}-\tau _{0}$
is orthogonal to the face $F\left( \rho _{0},\tau _{0}\right) $, and from
Corollary 5.3 we also know what the extreme points of $F\left( \rho
_{0},\tau _{0}\right) $ are. Now suppose that $\rho _{a}=\left| \psi
_{a}\right\rangle \left\langle \psi _{a}\right| $ where $\psi
_{a}=\sum_{k=0}^{d-1}a_{k}\left| kk\right\rangle $ with $0\leq a_{k}$, $%
\sum_{k}a_{k}^{2}=1$, and the $a_{k}$'s are close to $1/\sqrt{d}$. Then one
would expect that $\tau _{a}$ would also lie in $F\left( \rho _{0},\tau
_{0}\right) $ and thus, considered as vectors, that $\rho _{a}-\tau _{a}$
might be parallel to $\rho _{0}-\tau _{0}$. This could take the form 
\begin{equation}
\tau _{a}=\rho _{a}+t\left( \tau _{0}-\rho _{0}\right) 
\end{equation}
where $t=t\left( d,a\right) $ is a positive constant to be determined. This
particular representation works when $d=2$, and we obtain the same
constraints on the parameters $a_{0}$ and $a_{1}$ found earlier in \cite
{witru}. As a convention, we will assume that $a_{0}>$ $a_{1}$ throughout.

\begin{proposition}
In the case $d=2$, $\tau _{a}=\rho _{a}+t\left( \tau _{0}-\rho _{0}\right) $
lies in $F\left( \rho _{0},\tau _{0}\right) $ and is the closest separable
density to $\rho _{a}$ provided $t=2a_{0}a_{1}$ and $\left| a_{0}^{2}-\frac{1%
}{2}\right| \leq \frac{\sqrt{5}}{6}$. The Frobenius measure of entanglement
is then $m\left[ \rho _{a}\right] =2a_{0}a_{1}/\sqrt{3}=2a_{0}a_{1}m\left[
\rho _{0}\right] $.
\end{proposition}

\textit{Proof}: If $\tau _{a}$ were in $F\left( \rho _{0},\tau _{0}\right) $%
, then $\left\langle \left\langle \rho _{0}-\tau _{0},\tau _{a}-\tau
_{0}\right\rangle \right\rangle =0$ and for any $\tau \in S$ 
\[
\left\langle \left\langle \rho _{a}-\tau _{a},\tau -\tau _{a}\right\rangle
\right\rangle =t\left[ \left\langle \left\langle \rho _{0}-\tau _{0},\tau
-\tau _{0}\right\rangle \right\rangle -\left\langle \left\langle \rho
_{0}-\tau _{0},\tau _{a}-\tau _{0}\right\rangle \right\rangle \right] \leq
0, 
\]
confirming that $\tau _{a}$ would be the closest separable density to $\rho
_{a}.$ For 
\[
\tau _{a}=\left( 
\begin{array}{cccc}
a_{0}^{2}-t/6 & 0 & 0 & a_{0}a_{1}-t/3 \\ 
0 & t/6 & 0 & 0 \\ 
0 & 0 & t/6 & 0 \\ 
a_{0}a_{1}-t/3 & 0 & 0 & a_{1}^{2}-t/6
\end{array}
\right) 
\]
to be in $F\left( \rho _{0},\tau _{0}\right) $ it has to be a convex
combination of the extreme points of $F\left( \rho _{0},\tau _{0}\right) $.
If the entries of $\left| \beta \right\rangle $ are denoted by $%
r_{k}e^{i\theta \left( k\right) }$, then the $\left(
j_{1}j_{2},k_{1}k_{2}\right) $'th component of $\tau =\left| \beta
\right\rangle \left\langle \beta \right| \otimes \left| \bar{\beta}%
\right\rangle \left\langle \bar{\beta}\right| $ is 
\begin{equation}
r_{j_{j}}r_{j_{2}}r_{k_{1}}r_{k_{2}}e^{i\left( \theta \left( j_{1}\right)
-\theta \left( j_{2}\right) -\theta \left( k_{1}\right) +\theta \left(
k_{2}\right) \right) }.
\end{equation}
It follows that $\tau _{00,11}=\tau _{11,00}=\tau _{10,10}=\tau
_{01,01}=r_{0}^{2}r_{1}^{2}$ is always real, and it is easy to see that all
other entries have phase angles. Thus a necessary condition is $%
a_{0}a_{1}-t/3=t/6$, giving $t=2a_{0}a_{1}$ as asserted.

Keeping $r_{0}$ and $r_{1}$ fixed and averaging extreme points with phase
angles changed appropriately by angles of $\pi /2$ and $\pi $, one can
eliminate non-zero entries where phase angles appear and thus define a
convex subset $\hat{F}\left( \rho _{0},\tau _{0}\right) $ of $F\left( \rho
_{0},\tau _{0}\right) $ defined by densities of the form 
\[
\left( 
\begin{array}{cccc}
r_{0}^{4} & 0 & 0 & r_{0}^{2}r_{1}^{2} \\ 
0 & r_{0}^{2}r_{1}^{2} & 0 & 0 \\ 
0 & 0 & r_{0}^{2}r_{1}^{2} & 0 \\ 
r_{0}^{2}r_{1}^{2} & 0 & 0 & r_{1}^{4}
\end{array}
\right) . 
\]
A necessary and sufficient condition for $\tau _{a}$ to be in this convex
subset of $F\left( \rho _{0},\tau _{0}\right) $ is that $\left( a_{0}^{2}-%
\frac{a_{0}a_{1}}{3},a_{1}^{2}-\frac{a_{0}a_{1}}{3},\frac{2a_{0}a_{1}}{3}%
\right) $ should be in the convex hull of vectors $\left(
r_{0}^{4},r_{1}^{4},2r_{0}^{2}r_{1}^{2}\right) $ with $r_{0}^{2}+r_{1}^{2}=1$%
. If $x_{0}$ and $x_{1}$ denote $r_{0}^{2}$ and $r_{1}^{2}$ respectively,
then it is easy to check that an equivalent condition is that $\left(
a_{0}^{2},a_{0}^{2}-\frac{a_{0}a_{1}}{3}\right) $ should be in the convex
hull of vectors $\left( x_{0},x_{0}^{2}\right) $ where $0\leq x_{0}\leq 1$.
But that set is precisely the set of pairs $\left( x,y\right) $ with $0\leq
x\leq 1$ and $x^{2}\leq y\leq x$. That means $a_{0}^{4}\leq a_{0}^{2}-\frac{%
a_{0}a_{1}}{3}$ or $1\leq 3a_{0}a_{1}$, which is equivalent to the condition
asserted in the statement of the proposition. (Since $d=2$, the same result
could be obtained by using the required positive definiteness of $\tau _{a}$
and the Peres-Horodecki theorem.) The calculation of $m\left[ \rho
_{a}\right] $ is immediate, completing the proof. $\ \;\square $

\begin{corollary}
If $a_{0}^{2}=\frac{1}{2}+\frac{\sqrt{5}}{6}$, then $r_{0}^{2}=$ $\left( 
\frac{7+3\sqrt{5}}{18}\right) ^{1/2}$ and $\tau _{a}$ is on the boundary of $%
D$.\ $\square $
\end{corollary}

The significance of Corollary 7.1 turns out to be that for $a_{0}^{2}>\frac{1%
}{2}+\frac{\sqrt{5}}{6}$ the vector $\rho _{a}-\tau _{a}$ is in fact not
parallel to $\rho _{0}-\tau _{0}$, and the techniques above do not give $%
\tau _{a}$. This same problem arose in \cite{witru} where it was conjectured
that $\tau _{a}$ could be computed using the root of a cubic polynomial.
Geometrically, that cubic is based on the assumption that the nearest
separable density to $\rho _{a}$ is on the boundary of $\hat{F}\left( \rho
_{0},\tau _{0}\right) $, which is reasonable since we have already seen that 
$\rho _{a}$ is separated from $\mathit{S}$ by the hyperplane containing $%
F\left( \rho _{0},\tau _{0}\right) $. The gap in the argument is that one
needs to show that $\tau _{a}$ has to be in $\hat{F}\left( \rho _{0},\tau
_{0}\right) $.

\smallskip The same approach works for $d\geq 3$ but with the need for
additional parameters. The entries of the extreme points $\left| \beta
\right\rangle \left\langle \beta \right| \otimes \left| \bar{\beta}%
\right\rangle \left\langle \bar{\beta}\right| $ have the same sort of
pattern as in the $d=2$ case with real entries $r_{j}^{2}r_{k}^{2}$ in
positions $\left( jj,kk\right) $, $\left( jk,jk\right) $, $\left(
kj,kj\right) $, and $\left( kk.jj\right) $ for $0\leq j,k\leq d-1$.This
means there are ${d \choose 2}$ sets of four entries for which the components
of any density in the convex hull of the extreme points must be constant,
and the single parameter $t$ in equation (5.1) doesn't suffice.

To obtain more parameters, we look for additional Hermitian matrices
orthogonal to $F\left( \rho _{0},\tau _{0}\right) $. In particular, define
the ${d \choose 2}$matrices $M_{jk}$, $0\leq j<k<d$, whose entries are $+1$
at $\left( jk,jk\right) $ and $\left( kj,kj\right) $, $-1$ at entries $%
\left( jj,kk\right) $ and $\left( kk,jj\right) $ and are $0$ elsewhere. Then
it is easy to check that for extreme points $\tau $ in $F\left( \rho
_{0},\tau _{0}\right) $%
\begin{equation}
Tr\left( M_{jk}\tau \right) =\tau _{jk,jk}+\tau _{kj,kj}-\tau _{jj.kk}-\tau
_{kk,jj}=0,
\end{equation}
and thus each Hermitian $M_{jk}$ is orthogonal to densities in $F\left( \rho
_{0},\tau _{0}\right) $. Use $I$ to denote the $d^{2}\times d^{2}$ identity
and note that 
\begin{equation}
\sum_{j<k}M_{jk}=I-d\rho _{0}.
\end{equation}
As noted in the preceding section, up to a multiplicative constant $I-d\rho
_{0}$ is the optimal entanglement witness $A_{0}$ based on $\rho _{0}$.

We break the analysis into two parts, first showing that the $\tau _{a}$
defined below is the closest separable density to $\rho _{a}$, \textit{%
assuming} $\tau _{a}$ is in $F\left( \rho _{0},\tau _{0}\right) $, and then
obtaining sufficient conditions on the $a_{k}$'s for $\tau _{a}$ to be in $%
F\left( \rho _{0},\tau _{0}\right) $. Recall that $\rho _{a}$ has entries $%
a_{j}a_{k}$ in positions $\left( jj,kk\right) $ and $\left( kk,jj\right) $
for $0\leq j,k\leq d-1$, so that $\sum_{k}a_{k}^{2}=1$. As a convention we
assume that $a_{0}\geq a_{1}\geq \ldots \geq a_{d-1}$. Set $%
a*a=\sum_{j<k}a_{j}a_{k}$ and impose the first constraint on the $a_{k}$'s: 
\begin{equation}
a_{d-1}^{2}\geq \frac{2a*a}{d\left( d+1\right) }.
\end{equation}

\begin{proposition}
Let $\left\{ M_{jk}\right\} $ be defined as above and let $M_{a}$ denote $%
\sum_{j<k}u_{jk}M_{jk}$. Define 
\begin{equation}
\tau _{a}=\rho _{a}+t\left( \tau _{0}-\rho _{0}\right) +M_{a},
\end{equation}
where $t=\frac{2a*a}{d-1}$ and $u_{jk}=\frac{1}{2}\left( a_{j}a_{k}-\frac{t}{%
d}\right) $. Then $\sum_{j<k}u_{jk}=0$, and $\tau _{a}$ is a trace one,
Hermitian matrix with non-negative entries on the diagonal. Moreover, $\rho
_{a}-\tau _{a}$ is orthogonal to $F\left( \rho _{0},\tau _{0}\right) $. 
\textit{If} in addition $\tau _{a}$ is in $F\left( \rho _{0},\tau
_{0}\right) $, then it is the closest separable density to $\rho _{a}$ and 
\begin{equation}
m\left[ \rho _{a}\right] =\sqrt{Tr\left[ \left( \rho _{a}-\tau _{a}\right)
^{2}\right] }=\sqrt{t^{2}\left( 1-\frac{2}{1+d}\right) +\sum_{j<k}4u_{jk}^{2}%
}.
\end{equation}
\end{proposition}

\textit{Proof}: Using the definitions we first compute the non-zero entries
of $\tau _{a}$: 
\begin{eqnarray*}
\tau _{a}\left( ii,ii\right)  &=&a_{i}^{2}+\frac{td}{1+d}\left( \frac{1}{%
d^{2}}-\frac{1}{d}\right) =a_{i}^{2}-\frac{2a*a}{d\left( d+1\right) } \\
\tau _{a}\left( jk,jk\right)  &=&\frac{td}{1+d}\frac{1}{d^{2}}+u_{jk}=\frac{1%
}{2}\left( a_{j}a_{k}-\frac{2a*a}{d\left( d+1\right) }\right) ,j\neq k, \\
\tau _{a}\left( jj,kk\right)  &=&a_{j}a_{k}-\frac{td}{1+d}\frac{1}{d}-u_{jk}=%
\frac{1}{2}\left( a_{j}a_{k}-\frac{2a*a}{d\left( d+1\right) }\right) ,j\neq %
k.
\end{eqnarray*}
By virtue of equation (7.5) $\tau _{a}$ has non-negative entries on the
diagonal and the desired pattern of values on the remaining entries. Note
that these same formulas work for $d=2$. One can show $Tr\left( \tau
_{a}\right) =1$ directly or by confirming that 
\[
\sum_{j<k}u_{jk}=\frac{1}{2}\left( a*a-{d \choose 2}\frac{t}{d}\right) =0
\]
so that $Tr\left( M_{a}\right) =0$. Since $Tr\left( M_{jk}\left( \rho
_{0}-\tau _{0}\right) \right) =\frac{d}{1+d}Tr\left( M_{jk}\left( \rho
_{0}-D_{0}\right) \right) =\frac{-2}{d}$, 
\[
Tr\left( M_{a}\left( \rho _{0}-\tau _{0}\right) \right) =\frac{-2}{d}%
\sum_{j<k}u_{jk}=0.
\]
Thus, as a vector $\rho _{a}-\tau _{a}$ can be viewed as the sum of two
orthogonal vectors in $\left( F\left( \rho _{0},\tau _{0}\right) \right)
^{\bot }$, the linear subspace of Hermitian matrices perpendicular to $%
F\left( \rho _{0},\tau _{0}\right) $, and (7.7) follows from that
observation.

So far we have only shown that $\tau _{a}$ could be in $F\left( \rho
_{0},\tau _{0}\right) $, with equation (7.5) the only constraint imposed so
far on the $a_{k}$'s. To complete the proof of the proposition, we show that 
\textit{if} $\tau _{a}$ is in $F\left( \rho _{0},\tau _{0}\right) $ then it
is the nearest separable density to $\rho _{a}$. Since $\left\langle
\left\langle \rho _{0}-\tau _{0},\tau _{a}-\tau _{0}\right\rangle
\right\rangle =\left\langle \left\langle M_{jk},\tau _{a}\right\rangle
\right\rangle =0$ under that hypothesis, 
\begin{eqnarray*}
\left\langle \left\langle \rho _{a}-\tau _{a},\tau -\tau _{a}\right\rangle
\right\rangle &=&t\left[ \left\langle \left\langle \rho _{0}-\tau _{0},\tau
-\tau _{a}\right\rangle \right\rangle \right] -\sum_{j<k}u_{jk}\left\langle
\left\langle M_{jk},\tau -\tau _{a}\right\rangle \right\rangle \\
&=&t\left[ \left\langle \left\langle \rho _{0}-\tau _{0},\tau -\tau
_{0}\right\rangle \right\rangle \right] -\sum_{j<k}u_{jk}\left\langle
\left\langle M_{jk},\tau \right\rangle \right\rangle \\
&=&\frac{td}{d+1}\left\langle \left\langle \rho _{0}-D_{0},\tau -\tau
_{0}\right\rangle \right\rangle -\sum_{j<k}u_{jk}\left\langle \left\langle
M_{jk},\tau \right\rangle \right\rangle .
\end{eqnarray*}
Now it is easy to check that 
\[
\left\langle \left\langle \rho _{0}-D_{0},\tau -\tau _{0}\right\rangle
\right\rangle =\left\langle \left\langle \rho _{0},\tau \right\rangle
\right\rangle -\left\langle \left\langle \rho _{0},\tau _{0}\right\rangle
\right\rangle =\left\langle \left\langle \rho _{0},\tau \right\rangle
\right\rangle -\frac{1}{d}=\frac{1}{d}\left\langle \left\langle d\rho
_{0}-I,\tau \right\rangle \right\rangle , 
\]
where $I$ denotes the $d^{2}\times d^{2}$ identity matrix. Using $d\rho
_{0}-I=-\sum_{j<k}M_{jk}$ from equation (7.4) and 
\[
\sum_{j<k}u_{jk}\left\langle \left\langle M_{jk},\tau \right\rangle
\right\rangle =\frac{1}{2}\sum_{j<k}a_{j}a_{k}\left\langle \left\langle
M_{jk},\tau \right\rangle \right\rangle -\frac{t}{2d}\sum_{j<k}\left\langle
\left\langle M_{jk},\tau \right\rangle \right\rangle , 
\]
we have 
\[
\left\langle \left\langle \rho _{a}-\tau _{a},\tau -\tau _{a}\right\rangle
\right\rangle =-\frac{1}{2}\sum_{j<k}\left\langle \left\langle M_{jk},\tau
\right\rangle \right\rangle \left( \frac{t\left( d-1\right) }{d\left(
d+1\right) }+a_{j}a_{k}\right) . 
\]
We have already shown that $\left\langle \left\langle M_{jk},\tau
\right\rangle \right\rangle =0$ when $\tau $ is in $F\left( \rho _{0},\tau
_{0}\right) $. If $\tau $ is a separable extreme point of the form $\left|
\beta \right\rangle \left\langle \beta \right| \otimes \left| \gamma
\right\rangle \left\langle \gamma \right| $, then using the obvious notation
and equation (7.3) 
\begin{eqnarray*}
\left\langle \left\langle M_{jk},\tau \right\rangle \right\rangle &=&\tau
_{jk,jk}+\tau _{kj,kj}-\tau _{jj.kk}-\tau _{kk,jj} \\
&=&b_{j}^{2}c_{k}^{2}+b_{k}^{2}c_{j}^{2}-2b_{j}b_{k}c_{j}c_{k}\cos \left(
\theta _{jk}\right) \geq 0.
\end{eqnarray*}
Hence $\left\langle \left\langle \rho _{a}-\tau _{a},\tau -\tau
_{a}\right\rangle \right\rangle \leq 0$ for all separable densities, and
that completes the proof.\ \ $\square $

\smallskip As we saw in the case when $d=2$, further restrictions on the $%
a_{k}$'s are required to show that $\tau _{a}$ actually is a separable
density. We examine that problem next and obtain sufficient conditions on
the $a_{k}$'s for $\tau _{a}$ to be in $F\left( \rho _{0},\tau _{0}\right) $%
. It then follows from the foregoing analysis that $\tau _{a}$ is the
separable density closest to the related density $\rho _{a}$, confirming the
intuition that motivated this analysis in the first place. Unfortunately the
algebra appears to be too involved to get as precise a result as in the case
when $d=2.$

What we do instead is demonstrate a methodology which shows that there
exists a neighborhood of the equal entry case when $a_{k}=1/\sqrt{d}$ in
which $\tau _{a}$ is in $F\left( \rho _{0},\tau _{0}\right) $. We have
already shown that $\tau _{a}$ could be in the smaller convex set $\hat{F}%
\left( \rho _{0},\tau _{0}\right) $, and we follow the approach used in the $%
d=2$ case. $\tau _{a}$ will be in $\hat{F}\left( \rho _{0},\tau _{0}\right) $
if the ${d+1 \choose 2}$ vectors $\mathit{T}\left( a\right) $ whose first $d$
entries are $a_{i}^{2}-\frac{2a*a}{d\left( d+1\right) }$ and whose next $%
{d \choose 2}$ entries are $\left( a_{j}a_{k}-\frac{2a*a}{d\left( d+1\right) }%
\right) $, $j\neq k$, is in the convex hull of $\mathit{X}\left( x\right) $
vectors with respective entries $x_{i}^{2}$ and $2x_{i_{1}}x_{i_{2}}$, where 
$\sum_{i}x_{i}=1$ and $0\leq x_{i}$. Note that the components of all of the
vectors in question sum to $1$. In this notation $\tau _{0}$ corresponds to
a $\mathit{T}$-vector $\mathit{T}\left( 0\right) $ with entries $\frac{2}{%
d\left( d+1\right) }$.

The idea is to select a specific set of extreme $\mathit{X}$ vectors and
show that $\mathit{T}\left( 0\right) $ is in the interior of the convex hull
of these particular vectors. Specifically, for each of $1\leq k\leq d$ we
choose ${d \choose k}$ vectors $\mathit{X}\left( k;j\right) $, $1\leq j\leq 
{d \choose k}$, corresponding to a choice of $k$ of the $x_{i}$'s equal to $%
1/k$ and the remainder equal to $0.$ Thus, $\mathit{X}\left( k;j\right) $
will have $k$ of its entries corresponding to $x_{i}^{2}$ equal to $1/k^{2}$
and ${k \choose 2}$ entries corresponding to $2x_{i_{1}}x_{i_{2}}$ equal to $%
2/k^{2}$ . For example, if $d=3$ the resulting $7$ vectors can be written as
column vectors in a $6\times 7$ array $V$, and the assertion that $\mathit{T}%
\left( 0\right) $ is in the convex hull of these vectors is equivalent to $V%
\vec{p}=\mathit{T}\left( 0\right) $ or 
\begin{equation}
\left( 
\begin{array}{ccccccc}
1 & 0 & 0 & \frac{1}{4} & 0 & \frac{1}{4} & \frac{1}{9} \\ 
0 & 1 & 0 & \frac{1}{4} & \frac{1}{4} & 0 & \frac{1}{9} \\ 
0 & 0 & 1 & 0 & \frac{1}{4} & \frac{1}{4} & \frac{1}{9} \\ 
0 & 0 & 0 & \frac{1}{2} & 0 & 0 & \frac{2}{9} \\ 
0 & 0 & 0 & 0 & \frac{1}{2} & 0 & \frac{2}{9} \\ 
0 & 0 & 0 & 0 & 0 & \frac{1}{2} & \frac{2}{9}
\end{array}
\right) \cdot \left( 
\begin{array}{c}
p_{0} \\ 
p_{1} \\ 
p_{2} \\ 
p_{3} \\ 
p_{4} \\ 
p_{5} \\ 
p_{6}
\end{array}
\right) =\left( 
\begin{array}{c}
\frac{1}{6} \\ 
\frac{1}{6} \\ 
\frac{1}{6} \\ 
\frac{1}{6} \\ 
\frac{1}{6} \\ 
\frac{1}{6}
\end{array}
\right)
\end{equation}
with non-negative $p_{j}$'s summing to $1.$ Using this approach some easy
linear algebra shows that a sufficient condition for $\tau _{a}$ to be in $%
F\left( \rho _{0},\tau _{0}\right) $ when $a_{0}\geq a_{1}\geq a_{2}$ is
that 
\begin{equation}
\frac{1}{2}a_{0}a_{2}\leq a_{2}^{2}-\frac{1}{12}a*a\text{,}
\end{equation}
implying (7.5) when $d=3.$ As two examples, the $a$ corresponding to $a_{0}=%
\sqrt{5/12}$, $a_{1}=\sqrt{4/12}$, and $a_{2}=\sqrt{3/12}$ satisfies this
constraint, and the inequality (7.9) when $a_{k}=\frac{1}{\sqrt{3}}$ is $%
\frac{1}{6}\leq \frac{1}{4}$.

In the general case $V$ is a $\left( d+{d \choose 2}\right) \times \left(
\sum_{k=1}^{d} {d \choose k}\right) $ matrix with regular structure in each of
the $d$ blocks of ${d \choose k}$ columns. Each of the first $d$ rows will
have ${d-1 \choose k-1}$ non-zero entries equal to $1/k^{2}$ in the
corresponding block of ${d \choose k}$ columns. Similarly, each of the last $%
{d \choose 2}$ rows will have no positive entries in the first block of
columns and ${d-2 \choose k-2}$ non-zero entries equal to $2/k^{2}$ in the
remaining blocks of column vectors. If we further require that each of the $%
{d \choose k}$ column vectors have equal weight $q_{k}/{d \choose k}$, then
solving for $\vec{p}$ in $V\vec{p}=T\left( 0\right) $ is equivalent to
finding non-negative $q_{k}$ satisfying $\sum_{k}q_{k}=1$ and 
\[
\sum_{k=1}^{d}\frac{1}{k^{2}}q_{k}\frac{{d-1 \choose k-1}}{{d \choose k}}%
=\sum_{k=2}^{d}\frac{2}{k^{2}}q_{k}\frac{{d-2 \choose k-2}}{{d \choose k}}=%
\frac{2}{d\left( d+1\right) }.
\]
It is then easy to show that those three equations are equivalent to 
\begin{equation}
\sum_{k=1}^{d}q_{k}=1\hspace{0.5in}\sum_{k=2}^{d}\frac{q_{k}}{k}=\frac{2}{d+1%
}\text{.}
\end{equation}
Note that if $2k<d$, then $q_{k}=k/\left( d+1\right) $, $q_{d+1-k}=\left(
d+1-k\right) /(d+1)$ and $q_{j}$ equals zero otherwise is a particular
solution. From that observation it is easy to see that one can find
solutions of (7.10) which are strictly positive.

We have defined $2^{d}-1$ particular vectors $\mathit{X}$ which span their $%
{d+1 \choose 2}$ dimensional space. Given the components $a_{k}$ of $a$, we
want to find $\vec{p}_{a}$, a $\left( 2^{d}-1\right) $-long vector with
non-negative components which sum to one and such that $V\vec{p}_{a}=\mathit{%
T}\left( a\right) $, the vector corresponding to $\tau _{a}$. Equivalently,
we want to solve $V\vec{p}_{a}=V\vec{p}_{0}+V\left( \vec{p}_{a}-\vec{p}%
_{0}\right) =\mathit{T}\left( 0\right) +\left( \mathit{T}\left( a\right) -%
\mathit{T}\left( 0\right) \right) $\textit{. }By the earlier analysis we
know that we can choose a $\vec{p}_{0}$ whose entries are all strictly
positive, and thus the problem reduces to finding solutions of $V\vec{x}=%
\mathit{T}\left( a\right) -\mathit{T}\left( 0\right) $ where the components
of $\vec{x}$ sum to $0$ and are sufficiently small so that the components of 
$\vec{p}_{a}=\vec{p}_{0}+\vec{x}$ are non-negative. Since 
$2^{d}-1>{d+1 \choose 2}$, this is always possible, provided the components of $\mathit{T}\left(
a\right) -\mathit{T}\left( 0\right) $ are also sufficiently small. This
proves the final assertion of this section.

\begin{proposition}
If the components $a_{k}$ of $a$ are sufficiently close to $1/\sqrt{d}$,
then $\tau _{a}$ is in $F\left( \rho _{0},\tau _{0}\right) $.
\end{proposition}

\section{\textbf{Orthogonality in the n qubit case}}

In the absence of an efficient algorithm to compute the nearest separable
density to a given $\rho _{0}$ we have used the special structure of states
near maximally entangled states to find $\tau _{0}$. In particular we found
in Section 4 that in the bivariate case the nearest separable state to $\rho
_{0}\left( d\right) $ lay along the line in $\mathit{M}$ connecting $\rho
_{0}$ to $D_{0}$, and we also saw that was not true if there were more than
two systems. In this section we work with $n$ qubits and show that the
special structure of $\rho _{0}=\rho _{0}\left( n\right) =\left| \psi
_{0}\right\rangle \left\langle \psi _{0}\right| $ where $\left| \psi
_{0}\right\rangle =\frac{1}{\sqrt{2}}\left( \left| 00\ldots 0\right\rangle
+\left| 11\ldots 1\right\rangle \right) $ facilitates the analysis. In
particular we will obtain some perspective on the geometry in this higher
dimensional context.

The approach is straight-forward. We use the structure of $\rho _{0}$ as a
matrix in the computational basis and consider the local unitary mappings
which leave $\rho _{0}$ invariant. Since such operations should also leave $%
\tau _{0}$ invariant, we assume $\tau _{0}$ will have non-zero entries only
on the diagonal and in the $\left( \tilde{0},\tilde{1}\right) =\left(
00\ldots 0,11\ldots 1\right) $ and $\left( \tilde{1},\tilde{0}\right)
=\left( 11\ldots 1,00\ldots 0\right) $ positions. Additional considerations
of symmetry and positive definiteness reduce the calculation to a one
variable problem which can be solved by minimizing $\left\| \rho _{0}-\tau
_{0}\right\| $ over the remaining free parameter. The result of that
calculation provides a judicious guess for the form of $\tau _{0}$, and the
work is in the verification. These results include the two qubit case which
has $r_{2}=2$ in the notation below.

\begin{theorem}
For fixed $n\geq 2$ let $r_{n}=2^{n-1}$ and let $\tau _{0}$ denote the $%
2^{n}\times 2^{n}$ matrix with entries equal to $0$ except for 
\begin{eqnarray*}
\tau _{0}\left( \tilde{0},\tilde{0}\right) &=&\tau _{0}\left( \tilde{1},%
\tilde{1}\right) =a_{n}=\frac{r_{n}^{2}-2r_{n}+2}{2r_{n}^{2}-2r_{n}+2} \\
\tau _{0}\left( \tilde{0},\tilde{1}\right) &=&\tau _{0}\left( \tilde{1},%
\tilde{0}\right) =b_{n}=\frac{1}{2r_{n}^{2}-2r_{n}+2}
\end{eqnarray*}
and with all other entries on the diagonal also equal to $b_{n}$. Then 
\[
m\left[ \rho _{0}\right] =\left\| \rho _{0}-\tau _{0}\right\| =\frac{1}{%
\sqrt{2}}\left( 1-\frac{1}{r_{n}^{2}-r_{n}+1}\right) ^{\frac{1}{2}} 
\]
The extreme points of $F\left( \rho _{0},\tau _{0}\right) $ consist of $%
\left| \tilde{0}\right\rangle \left\langle \tilde{0}\right| $, $\left| 
\tilde{1}\right\rangle \left\langle \tilde{1}\right| $, and projections of
the form $\tau =\otimes _{k=1}^{n}\left| \psi _{k}\right\rangle \left\langle
\psi _{k}\right| $ where $\left| \psi _{k}\right\rangle =\frac{1}{\sqrt{2}}%
\left( e^{i\varphi _{k}/2}\left| 0\right\rangle +e^{-i\varphi _{k}/2}\left|
1\right\rangle \right) $ with $\Phi =\sum_{k}\varphi _{k}=0$ modulo $2\pi $.
\end{theorem}

\textit{Proof}: The calculation of $\left\| \rho _{0}-\tau _{0}\right\| $ is
routine, once we know that $\tau _{0}$ is the closest separable density.
Thus we want to show that $Tr\left( A_{0}\tau \right) \geq 0$ for separable $%
\tau $ when $A_{0}=c_{0}I+\tau _{0}-\rho _{0}$, and as usual it suffices to
check the inequality for separable projections. A routine calculation of $%
c_{0}=Tr\left( \tau _{0}\left( \rho _{0}-\tau _{0}\right) \right) $ gives $%
c_{0}=\frac{r_{n}-1}{2r_{n}^{2}-2r_{n}+2}$. A separable projection can be
written as the tensor product of $n$ matrices of the form 
\[
\left( 
\begin{array}{cc}
r_{k}^{2}\left( 0\right) & r_{k}\left( 0\right) r_{k}\left( 1\right)
e^{-i\varphi _{k}/2} \\ 
r_{k}\left( 0\right) r_{k}\left( 1\right) e^{i\varphi _{k}/2} & 
r_{k}^{2}\left( 1\right)
\end{array}
\right) , 
\]
and when we carry out the details we find that 
\[
Tr\left( A_{0}\tau \right) =\frac{r_{n}}{2r_{n}^{2}-2r_{n}+2}F\left( \tau
\right) 
\]
with 
\begin{equation}
F\left( \tau \right) =1-\prod_{k}r_{k}^{2}\left( 0\right)
-\prod_{k}r_{k}^{2}\left( 1\right) -\left( 2^{n}-2\right)
\prod_{k}r_{k}\left( 0\right) r_{k}\left( 1\right) \cos \left( \Phi \right)
\end{equation}
where $\Phi $ $=\sum_{k}\varphi _{k}$. Since $r_{k}^{2}\left( 0\right)
+r_{k}^{2}\left( 1\right) =1$, we can write the $1$ in $F\left( \tau \right) 
$ as the product of all $n$ terms $r_{k}^{2}\left( 0\right) +r_{k}^{2}\left(
1\right) $. Subtracting $\prod_{k}r_{k}^{2}\left( 0\right)
+\prod_{k}r_{k}^{2}\left( 1\right) $ from that product leaves $2^{n}-2$
terms of the form $\prod_{k}r_{k}^{2}\left( j_{k}\right) $ where the binary
indices $j_{k}$ are not all the same. These terms can be grouped in pairs so
that each factor of $r_{k}^{2}\left( 0\right) $ and $r_{k}^{2}\left(
1\right) $ appears in exactly one of the two paired terms. Then $F\left(
\tau \right) $ can be written as the sum of $2^{n-1}-1$ expressions of the
form 
\begin{equation}
\left[ \prod_{k}r_{k}^{2}\left( j_{k}\right) +\prod_{k}r_{k}^{2}\left( \bar{j%
}_{k}\right) -2\prod_{k}r_{k}\left( 0\right) r_{k}\left( 1\right) \cos
\left( \Phi \right) \right] ,
\end{equation}
where $\bar{j}_{k}$ denotes the binary complement of $j_{k}$. Since each of
these expressions is non-negative, $Tr\left( A_{0}\tau \right) \geq 0$ for
separable $S$.

Suppose $F\left( \tau \right) =0$ for $\tau $ a separable projection. Then
it's easy to check from equation (8.1) that if any one of the factors $%
r_{k}\left( 0\right) =1$, all of the factors $r_{j}\left( 0\right) =1$ and $%
\left| \tilde{0}\right\rangle \left\langle \tilde{0}\right| $ is in $F\left(
\rho _{0},\tau _{0}\right) $. Similar reasoning shows that $\left| \tilde{1}%
\right\rangle \left\langle \tilde{1}\right| $ is also in $F\left( \rho
_{0},\tau _{0}\right) $, and the only remaining case is when none of the
factors equals zero. Since each expression in equation (8.2) must be zero, $%
\cos \left( \Phi \right) =1$ and 
\[
\prod_{k}r_{k}\left( j_{k}\right) =\prod_{k}r_{k}\left( \bar{j}_{k}\right)
\neq 0 
\]
for all $n-$tuples $\left( j_{1},\ldots ,j_{n}\right) $. But then it is easy
to show that $r_{j}\left( 0\right) =r_{j}\left( 1\right) =1/\sqrt{2}$ for
all $j$, completing the characterization of the extreme points of $F\left(
\rho _{0},\tau _{0}\right) $ and the proof of the theorem.\ $\square $

\end{document}